\documentclass[12pt]{iopart}
\usepackage{iopams,graphicx} 
\usepackage{color}

\newcommand{\ket}[1]{\arrowvert #1 \rangle}

\begin{document}

\title{State-dependent interactions in ultracold $^{174}$Yb probed by optical clock spectroscopy}
\author{L. Franchi$^{1,2}$, L. F. Livi$^{3,2,4}$, G. Cappellini$^{1,2,4}$, G. Binella$^1$,\\M. Inguscio$^{4,1,3}$, J. Catani$^{4,2,3}$, L. Fallani$^{1,2,3,4}$}

\address{
$^1$Dipartimento di Fisica e Astronomia, Universit\`a degli Studi di Firenze, Via G. Sansone 1, I-50019 Sesto Fiorentino, Italy\\
$^2$INFN Istituto Nazionale di Fisica Nucleare, Sezione di Firenze, Via G. Sansone 1, I-50019 Sesto Fiorentino, Italy\\
$^3$LENS European Laboratory for Nonlinear Spectroscopy, Via N. Carrara 1, I-50019 Sesto Fiorentino, Italy\\
$^4$INO-CNR Istituto Nazionale di Ottica del CNR, Sezione di Sesto Fiorentino, Via N. Carrara 1, I-50019 Sesto Fiorentino, Italy

}
\ead{cappellini@lens.unifi.it}
\begin{abstract}
We report on the measurement of the scattering properties of ultracold $^{174}$Yb bosons in a three-dimensional (3D) optical lattice. Site occupancy in an atomic Mott insulator is resolved with high-precision spectroscopy on an ultranarrow, metrological optical clock transition. Scattering lengths and loss rate coefficients for $^{174}$Yb atoms in different collisional channels involving the ground state $^1$S$_0$ and the metastable $^3$P$_0$ states are derived. These studies set important constraints for future experimental studies of two-electron atoms for quantum-technological applications.
\end{abstract}

\submitto{\NJP}


\section{Introduction}

Ultracold neutral atoms are among the leading experimental platforms for the development of quantum technologies. Specifically, they offer vast opportunities for the engineering of synthetic many-body quantum systems, resulting in both the possibility to realize exact implementations of fundamental theoretical models -- in a quantum simulation perspective -- and to achieve new "extreme" states of matter \cite{qsim}. Experimental tools like optical lattices provide a prime method to control the atomic motion, laying a direct connection with solid-state physics \cite{lewenstein}. Remarkably, the interactions between the particles, and even new types of "synthetic" couplings with artificial fields, can be controlled by exploiting atomic-physics and quantum-optics techniques involving the manipulation of the internal atomic states.

Two-electron atoms, like alkaline-earth atoms and closed-shell rare-earth or lanthanide elements (such as ytterbium), feature a considerably large, accessible internal Hilbert space, with new advanced possibilities of quantum control, adding new tools to the quantum simulation \cite{gorkshov} and quantum information \cite{daley} toolboxes. Specifically, these atoms are characterized by long-lived electronic states (with lifetimes exceeding tens of seconds), that can be accessed by exciting the atoms on ultranarrow optical transitions. In particular, the optical clock transition connecting the ground state $g=$ $^1$S$_0$ to the long-lived state $e=$ $^3$P$_0$ has been used in the last two decades to engineer the best atomic clocks up-to-date, now outperforming conventional atomic clocks by several orders of magnitude \cite{reviewclocks1,reviewclocks2}.

Besides the metrological application, coherent manipulation of the electronic state opens new routes for the production of artificial magnetic fields \cite{gerbier} and, indeed, recently allowed for the demonstration of both spin-orbit coupling and tunable synthetic flux ladders \cite{spinorbit1,spinorbit2}, successfully extending the concept of synthetic dimensions \cite{celi,synthdim1,synthdim2} towards a new implementation. Recent experiments with fermionic strontium and ytterbium showed the possibility to take advantage from the interplay between the electronic degree of freedom and the nuclear spin, resulting in novel effects such as two-orbital spin-exchange \cite{spinex1,spinex2} and interaction tuning with a new kind of orbital Feshbach resonances \cite{ofr1,ofr2,ofr3}, possibly leading the way to novel states of matter \cite{gorkshov,iemini}. 

Most of these applications crucially rely on the scattering properties of atoms in different electronic states. While state-dependent scattering lengths for strontium \cite{traverso,zhang} and fermionic $^{173}$Yb \cite{spinex1,spinex2} have been thoroughly characterized, a similar study for the bosonic isotopes of ytterbium is still lacking. Knowing the state-dependent scattering properties of two-electron atoms is also highly relevant for the development of optical atomic clocks, especially for clocks based on bosonic isotopes, where low-temperature collisions cannot be prevented by the preparation of the atomic sample in the same initial internal state.

Here we are reporting on high-resolution clock spectroscopy of $^{174}$Yb bosons trapped in the lowest band of a 3D optical lattice in a Mott insulator state. These studies allow us to resolve the site occupancy in the Mott state and use this information to determine scattering lengths and loss rate coefficients for collisions involving atoms in the $e$ state, that were previously unknown.


\section{Experimental setup}

To address the clock transition we start preparing a Bose-Einstein condensate (BEC) of $ \sim 2\times10^5$  $^{174}$Yb atoms via evaporative cooling in a crossed optical dipole trap with trapping frequencies $\omega_{x,y,z}=2\pi\times(92.8,72.6,86.3)$ Hz. At the end of the evaporation stage the degenerate gas is loaded into a 3D cubic optical lattice operated at the magic  wavelength $\lambda_L=759.35$ nm \cite{Barber}, in order to have the same polarizability for the $g$ and the $e$ states. The lattice depth, measured in units of the lattice recoil energy $E_r=\hbar^2 k_L^2/2m$ (where $\hbar$ is the reduced Planck constant, $m$ is the atomic mass and $k_L=2\pi/\lambda_L$), can be tuned up to $s=40$. Once the lattice is loaded, the crossed optical trap is adiabatically turned off and the atoms remain trapped only by the lattice beams, in a Mott insulator state. After this process we end up with a sample consisting of  $\sim 1.2\times10^5$ atoms, a number that can be further controlled by changing the waiting time in the lattice after the optical trap has been turned off. 

\begin{figure}[tbp]
\centering
\includegraphics[width=.7\linewidth]{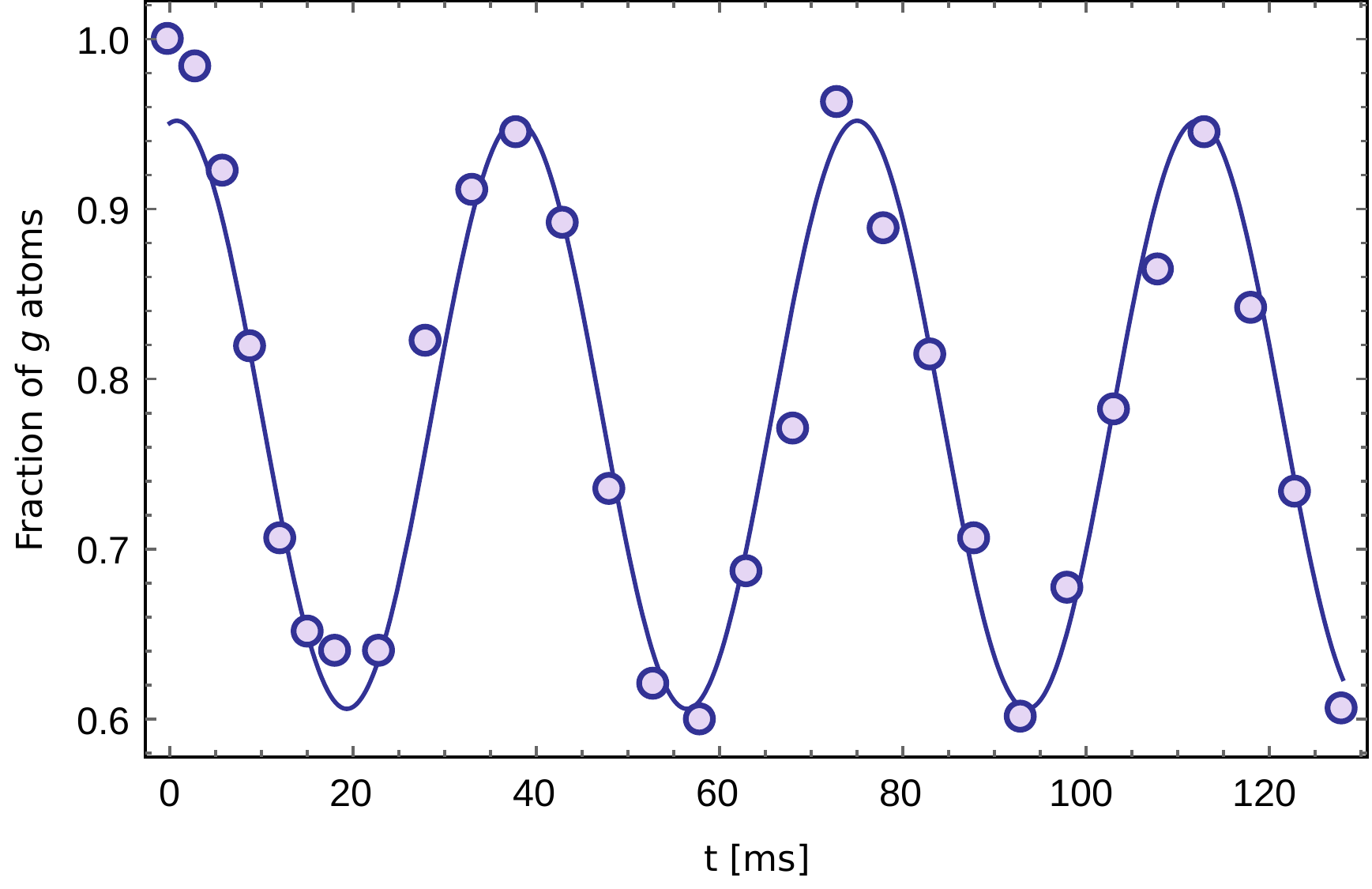}
\caption{Rabi oscillation after selective excitation of the singly-occupied sites of an atomic $^{174}$Yb Mott insulator in a deep 3D optical lattice. The number of $g$ atoms remaining after the clock excitation pulse is reported as a function of the pulse duration. The incomplete depletion that we observe at the $\pi$ pulse (after approx. 20 ms) can be explained by the presence of multiply-occupied lattice sites that are not probed in this measurement.}
\label{fig:Rabi}
\end{figure}

Spectroscopy is then performed illuminating the atoms with a 578 nm, $\pi$-polarized, laser pulse, resonant with the $\ket{g} \rightarrow \ket{e}$ clock transition. Being this transition strictly forbidden for bosons, the coupling is artificially induced \cite{taichenachev} by a magnetic field which ranges from 55 to 175 G, depending of the specific experiment. The excitation in the 3D lattice is performed in the Lamb-Dicke regime, with lattice depths spanning from $s=15$ to $s=40$. The linewidth of our clock laser \cite{cappellinilaser} is estimated to be a few tens of Hz on the timescale of the atomic excitation, as it allows to resolve spectroscopic features with a frequency width of a few tens of Hz. Figure \ref{fig:Rabi} shows a Rabi oscillation of the atomic population, showing a coherence time in the atom-laser interaction clearly exceeding $100$ ms. As the overall spectrum is recorded on a timescale of about 20 minutes, from the same data we can obtain information on the long-term stability of the laser during this measurement, with an estimated drift $<2$ Hz/min.

\begin{figure}[htbp]
\centering
\includegraphics[width=1\linewidth]{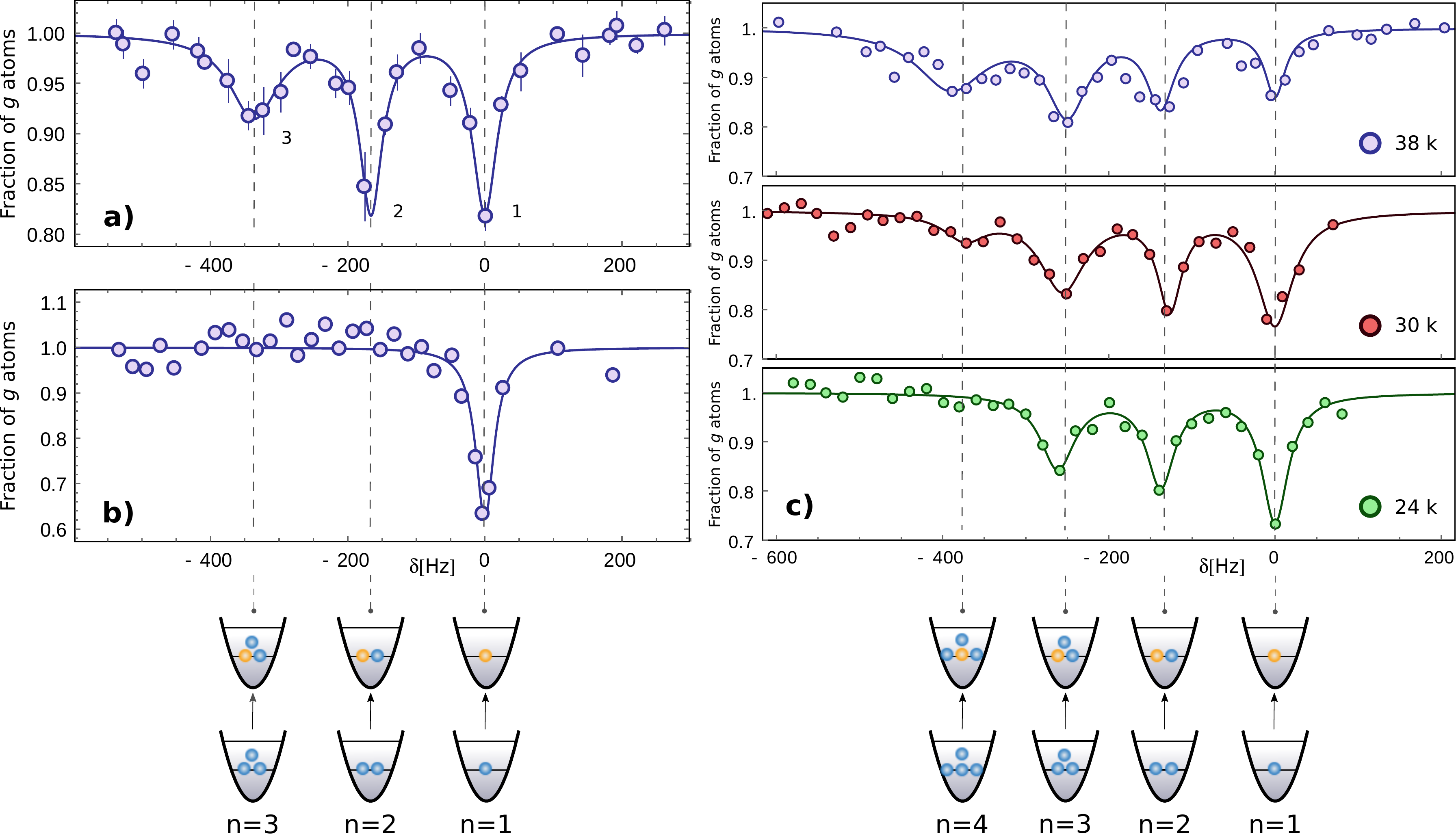}
\caption{Clock spectroscopy of an $^{174}$Yb atomic sample in a 3D optical lattice. All the graphs report the fraction of $g$ atoms remaining after the clock excitation pulse. \textbf{a}) In a typical spectrum we observe several evenly-spaced resonances, that we attribute to an interaction energy shift depending on the number of particles contained in each lattice site. The origin of the frequency axis has been chosen to match the position of the highest-energy resonance. \textbf{b}) Clock spectroscopy performed after the application of a photoassociation (PA) pulse, which removes the atoms in multiply-occupied sites, allowing the unambiguous identification of the resonance in singly-occupied sites, corresponding to the unshifted $g-e$ transition frequency. \textbf{c)} Clock spectroscopy performed for different numbers of atoms contained in the sample. As the atom number is decreased, the relative weight of the lower-energy resonances decreases, in agreement with the identification of the highest-energy resonance with the excitation of singly-occupied sites.}
\label{fig:spectroscopy}
\end{figure}


\section{Experimental spectra and determination of state-dependent scattering lengths}

In order to acquire the spectra, the atoms are probed with long interrogation pulses extending from 500 ms to 1.5 s, depending on the specific experiment. The number of atoms remaining in the ground state is then recorded as a function of the laser frequency, causing the spectroscopic signal to appear as a decrease in the atom number. Figure \ref{fig:spectroscopy}a reports a typical spectrum acquired for a sample of $\sim 5\times10^4$ atoms trapped in a 3D optical lattice with $s=30$. The spectrum is characterized by several resonances that we ascribe to processes where a single $g$ atom in a lattice site occupied by $n$ particles is excited to the $e$ state, a transition that we represent with the $\ket{(n)g}\rightarrow\ket{(n-1)g\,e}$ notation. We identify the highest-energy resonance in the spectra as the excitation of the $\ket{g}\rightarrow\ket{e}$ process in singly-occupied sites. This attribution is justified by the results of two different experiments in which we probe the transition by eliminating multiply-occupied sites in the lattice and by changing the total number of atoms in the sample, respectively.

In a first experiment (figure \ref{fig:spectroscopy}b) a photoassociation (PA) pulse is shone before the clock spectroscopy, leaving in this way a sample containing only singly-occupied sites. PA is performed using a 5-ms-long pulse having a frequency red-detuned by 20 MHz with respect to the MOT frequency \cite{photoassociation} and an intensity of $3\times 10^{-3}$ W mm$^{-2}$. No other resonances with the exception of the highest-energy one are observable in the resulting spectrum, which demonstrates the validity of our hypothesis.

In a second experiment (figure \ref{fig:spectroscopy}c) we perform spectroscopy on samples containing different atom numbers. As the atom number is lowered (and the mean number of particles in each lattice site is reduced) the relative weight of the lower-energy resonances decreases, which indicates that these processes can be attributed to $\ket{(n)g}\rightarrow\ket{(n-1)g\,e}$ transitions with $n>1$. On the other hand, the weight of the highest-energy resonance increases as the atom number is reduced, which signals an increased relative number of singly-occupied sites, further confirming the validity of our hypothesis. In additional Rabi oscillation measurements, we have validated our attribution of the particle numbers associated to the $n>1$ resonances (sketched in figure \ref{fig:spectroscopy}), by verifying that the Rabi frequency associated to the different resonances scales with the expected behavior $\Omega_n = \sqrt{n} \Omega_1$ \cite{dicke}.

These results indicate that it is possible to use the optical clock transition of Yb to perform an accurate spectroscopy of an atomic Mott insulator state, with a similar approach to those reported in Refs. \cite{Mottketterle,Mottjapan} by using different radiofrequency or optical transitions.

\subsection{Measurement of e-g scattering length}
\label{section_age}

\begin{figure}
\centering
\includegraphics[width=0.8\linewidth]{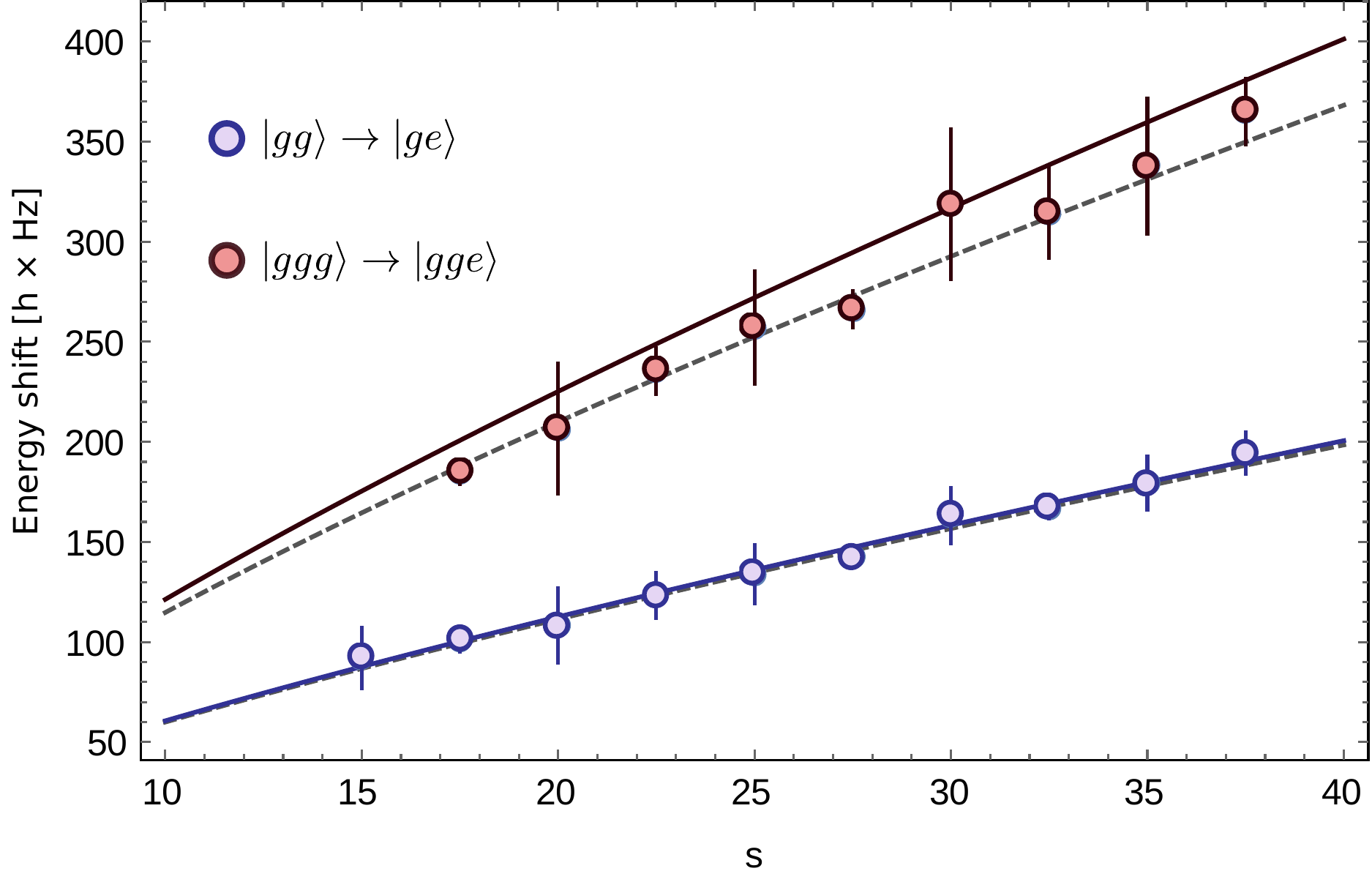}
\caption{Interaction shifts of the clock transition frequency for $n=2$ (blue) and $n=3$ (red) interacting $^{174}$Yb atoms, relative to the single-particle excitation frequency. The solid blue curve is a fit of the $n=2$ experimental data with equation \ref{delta_U2}. The solid red curve is the calculated shift for $n=3$ assuming the results of the previous fit to the $n=2$ data and taking into account only two-body interactions. The dashed curve is a simultaneous fit of the two experimental datasets taking into account, in an effective way, the energy correction for three-body elastic interactions reported in Ref. \cite{Johnson2009} (see text for more details).}
\label{fig:interaction_energy}
\end{figure}

The identification of the various resonances observed in figure \ref{fig:spectroscopy}a represents the starting point for the determination of the $s-$wave scattering length $a_{eg}$ in $^{174}$Yb, whose value has still not been reported, to our knowledge. We determine this quantity by measuring the interaction shift $\Delta U_{eg}=U_{eg}-U_{gg}$, that can be obtained from a fit of the experimental spectra as the energy difference between the centers of the $n=2$ and of the $n=1$ resonances, respectively labeled as (2) and (1) in figure \ref{fig:spectroscopy}a.
In the framework of the Bose-Hubbard model, $\Delta U_{eg}$ is related to the scattering length $a_{eg}$ by the following relation:
\begin{equation}
\label{delta_U2}
\Delta U_{eg}=\frac{4\pi\hbar^2  } {m} \,( a_{eg}-a_{gg} )\, \int w^4(\mathbf{r})\,d\mathbf{r} \, ,
\end{equation} 
where $a_{gg}$ is the scattering length for two ground-state atoms and $w(\mathbf{r})$ are the Wannier functions. The values of $\Delta U_{eg}$ measured for several depths of the 3D lattice are reported in figure \ref{fig:interaction_energy} (blue points). The experimental data have been fitted with equation \ref{delta_U2}, leaving the differential scattering length $a_{eg}-a_{gg}$ as the only free parameter. The result of the fit (solid line) corresponds to a best-fitting value
\begin{equation}
a_{eg}-a_{gg}= -10.19\,(0.13)\,a_0 \,.
\end{equation}
Combining this determination with the known value of $a_{gg}=+104.9\,(1.5)\,a_0$ reported in Ref. \cite{kitagawa2008} results in a value for the $g-e$ scattering length
\begin{equation}
a_{eg}= +94.7\,(1.6)\,a_0 \,.
\end{equation}

In figure \ref{fig:interaction_energy} we also plot the interaction shift of the $n=3$ resonance (labeled as (3) in figure \ref{fig:spectroscopy}a) relative to the $n=1$ resonance (red points). If only two-body interactions are taken into account, the total interaction energy of three interacting bosons should be $3U_{gg}$ if the particles are all in the $g$ state and $U_{gg}+2U_{eg}$ if one atom is excited to the $e$ state. This means that we expect the interaction energy shift to be $2(U_{eg}-U_{gg})=2\Delta U_{eg}$, i.e. twice that measured for the $n=2$ case. However, the expected shift (solid red line in figure \ref{fig:interaction_energy}), evaluated on the basis of the previous determination of $a_{eg}$, clearly deviates from the experimental points. We ascribe this deviation to an additional energy correction arising from elastic three-body effective interaction. This correction was introduced in Ref. \cite{Johnson2009} for the case of $n$ indistinguishable interacting bosons in the lowest vibrational state of a 3D optical lattice, and was observed experimentally in Refs. \cite{Johnson2009,ThreebodyBloch}. Following the derivation of \cite{Johnson2009}, for three particles in the same quantum state, at the first perturbative order, this correction reads

\begin{equation}
\label{delta_U3}
\delta U_{3}(a,s)= \frac{\beta \, U_{2}(a,s)^2}{\hbar \, \omega(s) / (2\pi)},
\end{equation} 
where $U_{2}(a,s)$ is the two-particle interaction energy, which depends on the scattering length $a$ and the lattice depth $s$, $\omega(s)$ is the harmonic frequency characterizing the confinement within one lattice site and $\beta=-1.34$ is a constant. This expression allows us to evaluate the correction to the interaction energy for the $\ket{ggg}$ state. For the $\ket{gge}$ state, however, the three interacting particles are not identical, and the theoretical approach of Ref. \cite{Johnson2009} could not be adequate. If we assume that the three-body elastic interaction in the $\ket{gge}$ state could still be described by equation \ref{delta_U3}, with an average scattering length given by the geometric mean $(a_{gg}  a_{eg}   a_{eg})^{1/3}$, also the measured $\ket{ggg}\rightarrow\ket{gge}$ interactions shifts can be used to estimate $a_{eg}$. In particular, a combined fit of the data relative to the two- and three-particles interaction shifts as a function of the lattice depths gives now a very good agreement with both datasets (dashed lines in figure \ref{fig:interaction_energy}). From the fit we extract the only fit parameter $\Delta a_{eg-gg} = a_{eg}-a_{gg}= -10.08\,(0.05)\,a_0$, which is consistent with the previous determination that does not rely on the knowledge of the elastic three-body contribution.

\subsection{Measurement of e-e scattering length}

\begin{figure}
\centering
\includegraphics[width=0.8\linewidth]{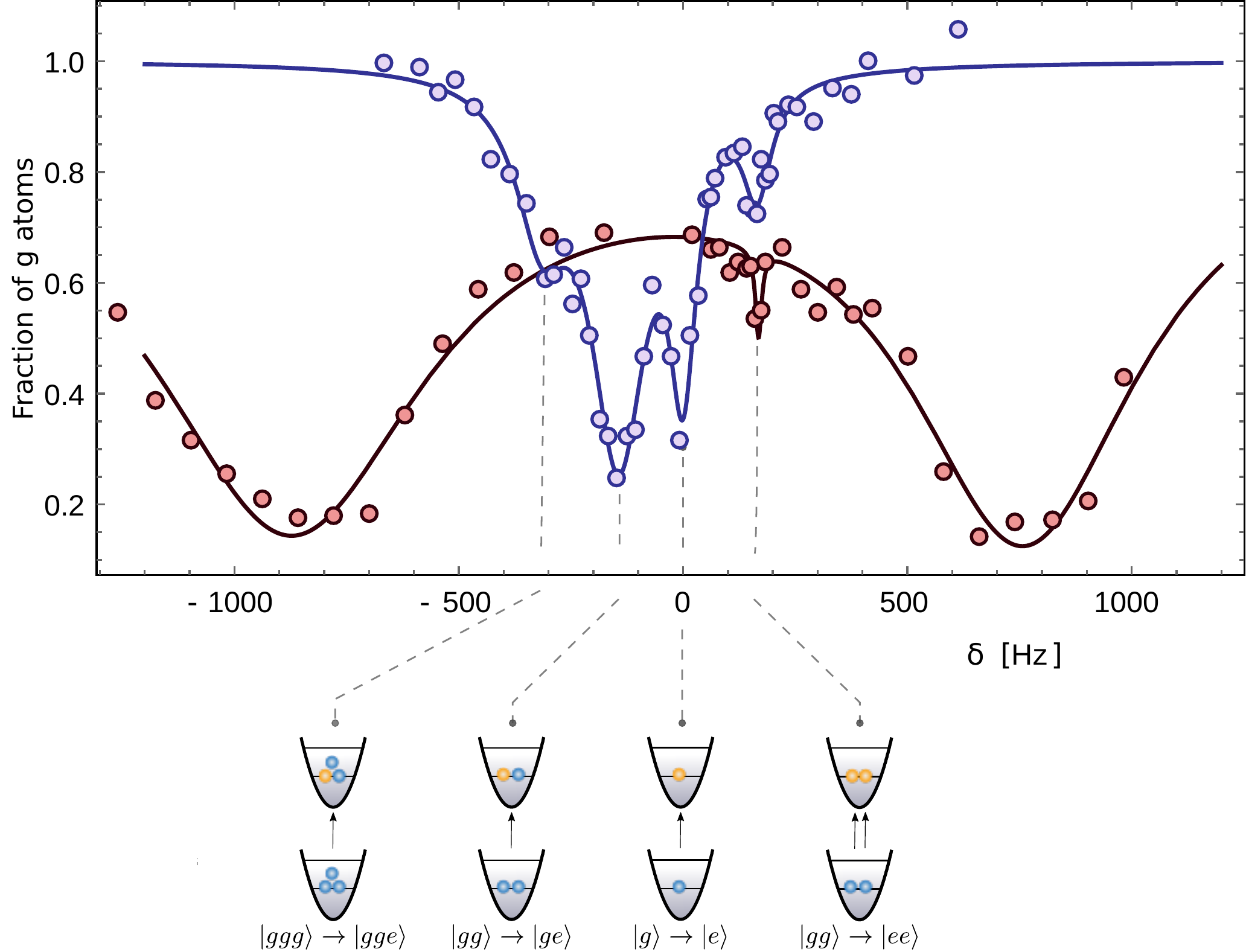}
\caption{Single-frequency (blue) and two-frequency (red) clock spectroscopy of an atomic $^{174}$Yb sample trapped in a 3D optical lattice with $s=30$. In the single-color spectrum, we identify the weak resonance at positive detuning as a signature of the two-particle/two-photon $\ket{gg}\rightarrow\ket{ee}$ process. This identification is confirmed by the presence of a resonance having the same frequency in the two-color spectrum taken at frequencies $f-\delta f$ and $f+\delta f$ (see text for more details). We note that the two-color spectrum has been recorded at a higher laser intensity in order to make the two-photon peak visible, causing the structure of the single-photon sidebands to be unresolved because of power broadening.}
\label{ee_twophotonspectroscopy}
\end{figure}

In our spectra, an increased coupling on the clock transition reveals the presence of an additional resonance that lies at a higher energy than the single-particle one, as shown in figure \ref{ee_twophotonspectroscopy} (blue points), at a frequency difference $\Delta f_{ee} \simeq 160$ Hz for $s \simeq 30$. We identify this resonance with the two-photon process $\ket{gg}\rightarrow\ket{ee}$, which transfers two particles trapped in the same lattice site from the $\ket{gg}$ to the $\ket{ee}$ state, via an intermediate $\ket{ge}$ state.

The two-photon nature of this excitation is validated by a two-frequency spectroscopy experiment in which the transition is excited by a clock laser with two frequency components $f_{+,-}=f \pm \delta f$ simultaneously probing the atomic sample.  The result of this experiment, performed fixing $\delta f = 800$ Hz and scanning $f$, is reported in figure \ref{ee_twophotonspectroscopy} (red points). The broad excitation profiles at the edges of the spectrum can be identified with two replicas of the single-photon absorption spectra, frequency-shifted by $+\delta f$ and $-\delta f$ respectively, as expected for a single-photon excitation driven by each of the two frequency components (the interactions sidebands are unresolved in this spectrum because of a larger laser intensity resulting in an increased power broadening). In addition, we still observe a weaker resonance at the same detuning $\Delta f_{ee} \simeq 160$ Hz as in the single-color spectrum, which is a strong indication of its two-photon nature. The absence of frequency shift for this resonance can be explained if we assume that a pair of atoms in the same lattice site absorbs simultaneously two photons with frequencies $f + \delta f$ and $f - \delta f$ each, in such a way that the total energy transferred to the system is $2 h f$ (where $h$ is the Planck constant), as in the case of a two-photon transition occurring for a single-color excitation.

The identification of the $\ket{gg} \rightarrow \ket{ee}$ resonance allows the experimental determination of the interaction energy shift $\Delta U_{ee} = U_{ee}-U_{gg}$ from the energy-conservation relation $2 h \Delta f_{ee} = \Delta U_{ee}$. Adopting an argumentation similar to the one carried on in section \ref{section_age}, it is possible to link this interaction shift to the scattering length $a_{ee}$ via the relation
\begin{equation}
\label{delta_Uee}
\Delta U_{ee}=\frac{4\pi\hbar^2  } {m} \,( a_{ee}-a_{gg} )\, \int w^4(\mathbf{r})\,d\mathbf{r} \, .
\end{equation} 
Using this equation and averaging over several spectra (both single-color and two-color) acquired at a mean lattice depth $s=29.3 (0.3)$, we determine the differential scattering length 
\begin{equation}
a_{ee} - a_{gg} = +21.8 \, (1.8) \, a_0 \, .
\end{equation}
Combining this measurement with the known value of $a_{gg}=+104.9\,(1.5)\,a_0$ reported in Ref. \cite{kitagawa2008} results in a value for the $ee$ scattering length
\begin{equation} 
a_{ee} = +126.7 \, (2.3) \, a_0 \, .
\end{equation}

\begin{table}
\caption{\label{tabone}Summary of the measured s-wave scattering lengths of $^{174}$Yb for different interaction channels involving the $g=$ $^1$S$_0$ and the $e=$ $^3$P$_0$ states (values in units of the Bohr radius $a_0$).} 
\begin{indented}
\lineup
\item[]\begin{tabular}{@{}*{3}{l}}
\br   
Scattering channel & Measured value & Reference \cr
\mr
$a_{gg}$& $+104.9\,(1.5)$ & \cite{kitagawa2008} \cr
$a_{eg}$& $+94.7\,(1.6)$ & This work + \cite{kitagawa2008}\cr 
$a_{ee}$& $+126.7 \, (2.3)$ & This work + \cite{kitagawa2008} \cr 
\mr                              
$a_{eg}-a_{gg}$& $-10.19\,(0.13)$
 & This work \cr 
$a_{ee} - a_{gg}$& $+21.8 \, (1.8) $
 & This work \cr 
\br
\end{tabular}
\end{indented}
\end{table}


\section{Detection of state-dependent inelastic collisions}

In this section we present the measurement of the loss rate coefficients relative to the $e-g$ and $e-e$ interaction channels. 

In these experiments, we detect the atoms in the metastable $^3$P$_0$ state ($e$) by repumping them to the $^3$D$_1$ state with a laser at 1389 nm. Atoms in the $^3$D$_1$ state can decay to any of the $^3$P$_J$ ($J=0,1,2$) states: while the atoms decayed to the $^3$P$_0$ state undergo another repumping cycle, atoms in the $^3$P$_1$ state decay to the ground state $^1$S$_0$ (in less than 1 $\mu$s), where they can be then detected. A small branching ratio limits the number of atoms lost into the metastable $^3$P$_2$ dark state, allowing for a high repumping efficiency of $e$-state atoms without any additional repumping laser. In our setup, we let the atoms interact with the 1389 nm light for a few ms during the time-of-flight, which grants a repumping efficiency $\gtrsim 90\%$. This is estimated by exciting an atomic sample in purely singly-occupied lattice sites to the $e$ state (with a $\pi-$pulse), then shining the repumping pulse and comparing the measured number of atoms to that of a non-excited $g$ state sample. Before the repumping pulse, the atoms in the ground state are blasted with a 100-$\mu$s-long pulse of imaging light during time of flight, so that only the $e$ state atoms are detected.

\subsection{Inelastic $e-g$ collisions}

\begin{figure}
\centering
\includegraphics[width=1\linewidth]{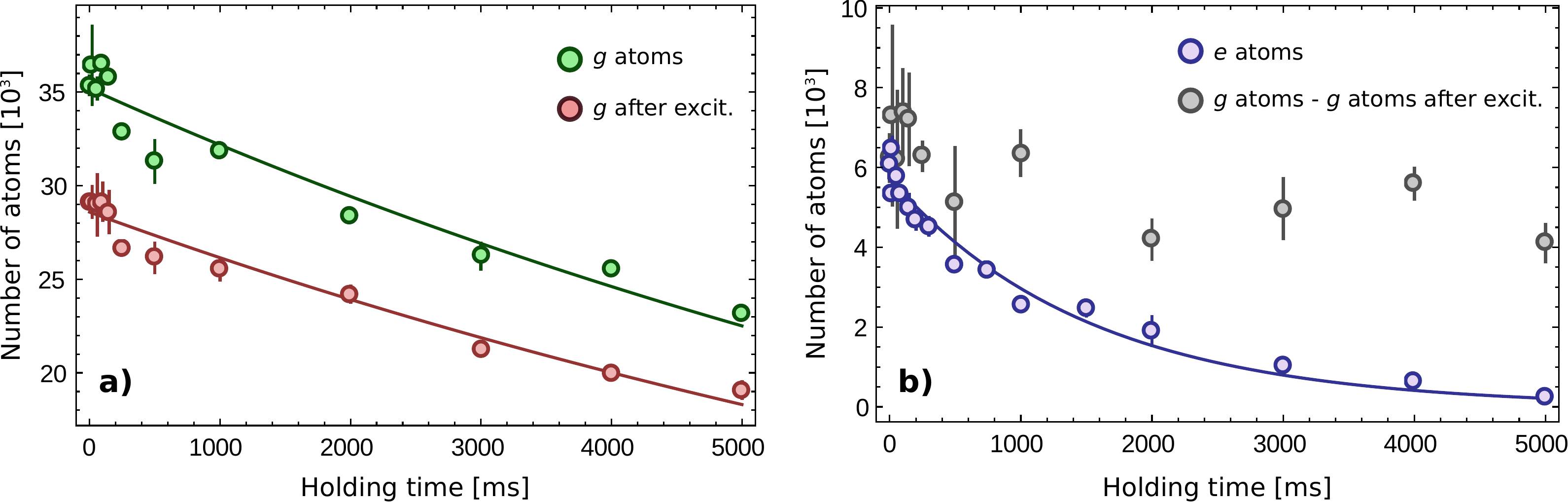}
\caption{Inelastic $e-g$ collisions have been investigated by exciting atoms in the $\ket{gg}$ state to the $\ket{eg}$ state by selectively addressing the $\ket{gg}\rightarrow\ket{eg}$ transition in a 3D optical lattice ($s=27.5$). \textbf{a}) The number of atoms remaining in $g$ after the clock pulse (red) is compared with the number of atoms in the absence of clock excitation  (green) as a function of the holding time in the lattice (the slow decay of the green points shows the finite single-particle lifetime in the lattice). Lines are exponential fits of the experimental data that must be intended as guides to the eye. \textbf{b}) The number of atoms detected in the $e$ state (blue) is shown as a function of the holding time in the lattice after the excitation of the transition. This number is compared with the difference between the number of $g$ atoms without and with the clock excitation (gray), which displays an approximately time-independent behavior. The solid blue line is an exponential fit of the experimental data.}
\label{fig:eg_decay}
\end{figure}

In a first experiment, we investigate the $e-g$ losses by loading the atomic sample in a 3D optical lattice with an average depth of s $\simeq$ 29.7. The waiting times in the trap (before the lattice loading) and in the lattice are adjusted in order to maximize the number of doubly-occupied sites and at the same time minimizing the number of multiply-occupied sites. The atoms in the $\ket{gg}$ state are then excited to the $\ket{eg}$ state by means of a 10-ms-long $\pi$-pulse of the clock laser selectively resonant with the $\ket{gg}\rightarrow\ket{eg}$ transition, and after a variable holding time in the lattice the number of atoms in the $e$ and in the $g$ states is measured. The collected data are reported in the two panels of figure \ref{fig:eg_decay} as blue points and red points for the $e$ and $g$ states respectively, while the green points represent the number of atoms in the $g$ state without performing the excitation pulse to the $e$ state. The solid lines are exponential fits to the data as guides to the eye. 

The data clearly show that atoms in the $e$ state decay on a timescale of the order of the second. To better understand this behavior, in figure \ref{fig:eg_decay}b we compare the number of atoms in the $e$ state (in blue) to the difference between the number of atoms in the $g$ state without and with the clock laser pulse (in gray). At short times, this difference is equal to the number of atoms in the $e$ state, as can be reasonably expected. On longer timescales of the order of $1$ s, instead, while the atoms in the $e$ state are lost, the difference between the $g$ atoms without and with the excitation is approximately constant, implying that the losses in the $e$ state are not caused by inelastic collisions with $g$ atoms, otherwise this difference should have increased blue(the small decrease could be attributed to the finite single-particle lifetime of the atoms in the lattice). The decay rate $\gamma = 1/\tau$ of the atoms in the $e$ state can be estimated with an exponential fit to the data, obtaining $0.66 (0.06)$ s$^{-1}$. This timescale is comparable with the tunneling rate $\sim 1$ Hz at the lattice depth of the experiment, suggesting that the $e$ atoms, initially in lattice sites with $g$ atoms, could be lost after tunnelling to neighboring sites via $e-e$ inelastic collisions. Since the timescale of the observed losses is determined by the tunnelling time before the actual interaction events, it is difficult to extract a reliable $e-e$ loss rate coefficient from those data.

Nevertheless, the observed dynamics allows us to give an upper bound to the $e-g$ inelastic loss rate coefficient. Two-body $e-g$ losses would be described by the rate equation $\dot{n}_g = -\beta_{eg} n_e n_g= -\gamma_{eg} n_g$, where $\beta_{eg}$ is the density-dependent loss rate coefficient. Requesting $\gamma_{eg} = \beta_{eg} n_e \ll \gamma$ and determining the in-site density $n_e$ from the calculated Wannier functions in the 3D lattice, we obtain
\begin{equation}
\beta_{eg} \ll 10^{-14} \; \mathrm{cm}^3/\mathrm{s} \; .
\end{equation}

\subsection{Inelastic $e-e$ collisions}
\begin{figure}
\centering
\includegraphics[width=.7\linewidth]{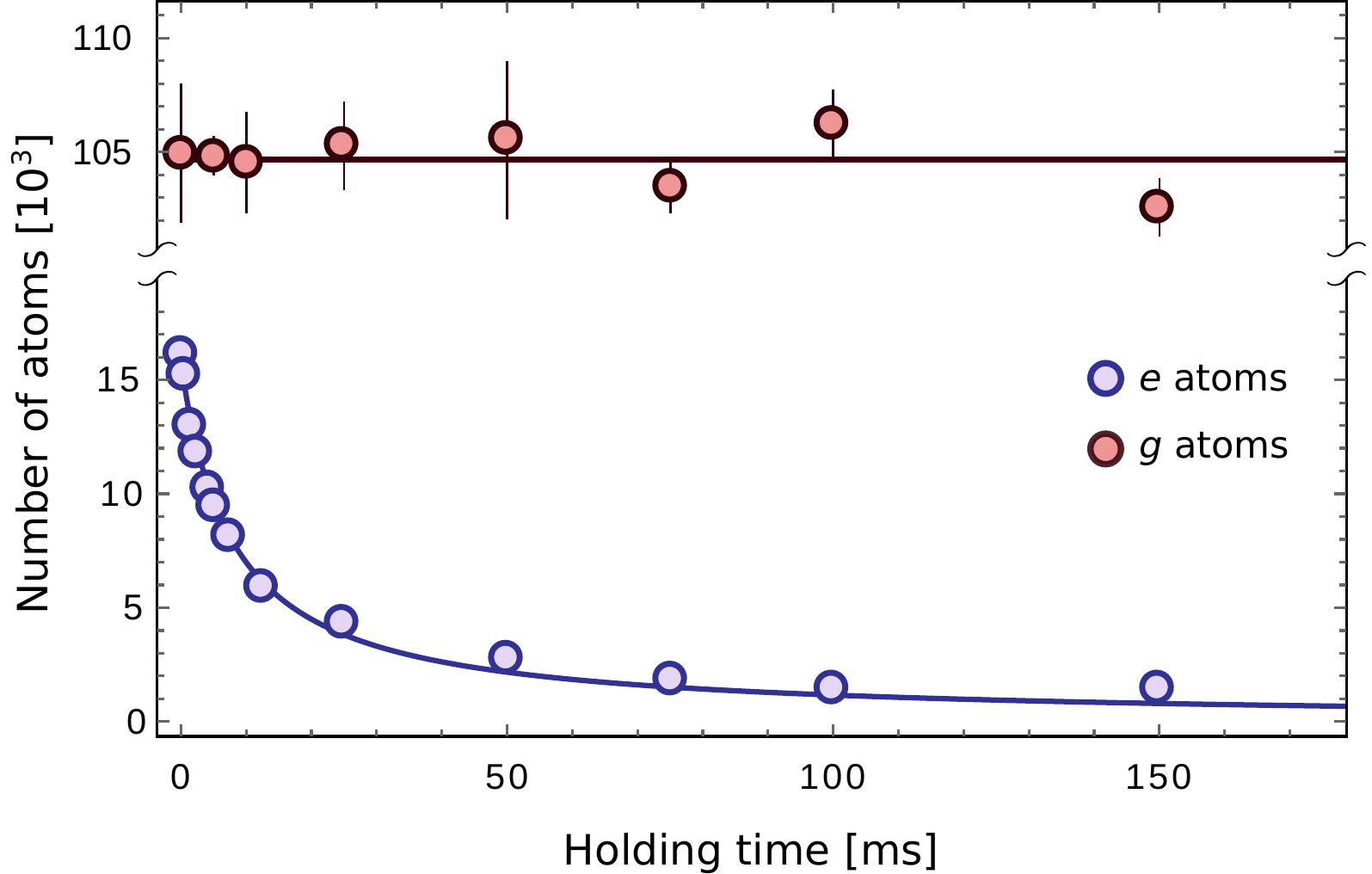}
\caption{
Number of $e$ atoms (in blue) and $g$ atoms (in red) as a function of the holding time in the lattice after a clock excitation pulse in an 1D lattice. The red line represents the mean value of $g$ atoms, while the blue line is the result of a fit to the $e$ data with the model in equation (\ref{eq:2bodymodel}), assuming the scaling in equation (\ref{eq:scalingnN}).}
\label{fig:ee_decay}
\end{figure}

A different strategy had to be implemented for the determination of the $e-e$ losses. As a matter of fact, we could not coherently excite a detectable number of atoms in the $ee$ state by means of $\pi$-pulses on the $\ket{gg}\rightarrow\ket{ee}$ transition, possibly due to an insufficient broadening of the two-photon transition. We then switched to a different geometry and loaded the atoms in a 1D vertical optical lattice at a depth $s=27.5$, obtaining an array of 2D pancakes with a radial trapping frequency $\omega = 2\pi \times 34.5$ Hz. A fraction of $g$ atoms is excited to the $e$ state with a 10-ms-long pulse of clock laser light directed along the pancakes plane, followed by a variable hold time in the lattice. Finally, the number of atoms in the $e$ and $g$ states is measured. As shown in figure \ref{fig:ee_decay}, the data exhibit fast losses in the $e$ state (blue points) on the ms timescale, while the $g$ population (red points) is constant (the red solid line represent the mean of the experimental data). On this timescale, as seen in the previous experiment, other loss channels are negligible, so, assuming only $e-e$ 2-body inelastic scattering (that is the dominant loss mechanism observed with different Yb and Sr isotopes, see e.g. Refs. \cite{traverso,spinex1}, and attributed to principal-number-changing collisions), the losses can be modeled with the rate equation $\dot{n}_e = -\beta_{ee} n_e^2$, where $n_e$ is the density of $e$ atoms and whose solution is given by:

\begin{equation} 
n_e(t) = \left( \frac{1}{n_{e0}} - \beta_{ee} t\right)^{-1} \; ,
\label{eq:2bodymodel}
\end{equation}
where $n_{e0}$ is the initial $e$ atom density and $\beta_{ee}$ is the 2-body loss rate coefficient. In order to extract a value for $\beta_{ee}$ from our data we developed a simplified theoretical model to determine the density in the pancakes from the measured atom number. 
Starting from the initial atom number $N_{g0} \simeq 180\times 10^3$ and density distribution of the sample in the optical dipole trap, we first determine the number of atoms in every pancake, assuming no population redistribution during the loading of the lattice. We then consider the density profile in the pancakes as a sum of a condensed part ($\sim 20\%$) and a thermal component at temperature $T_P \simeq 45$ nK, as determined from the experimental time-of-flight images integrated along the lattice axis. We also take into account the atom number reduction due to the 3 s wait time after loading the pancakes. We calculate the mean values of the density distribution in each pancake, and a final global average density $n_{g0}$ of the sample (before the clock excitation pulse) is determined by a weighted average over the pancake distribution. We then assume a linear relation between the atom number and the density (justified by the short timescale of the excitation), so that the $e$-state density $n_e(t)$ after the clock laser excitation can be determined as
\begin{equation}
n_{e}(t) = \frac{n_{g0}}{N_{g0}}N_{e}(t)
\label{eq:scalingnN}
\end{equation}
where $N_e(t)$ is the measured number of atoms in the $e$ state. Using this relation, we convert $N_e(t)$ into a density $n_e(t)$ and fit it with equation (\ref{eq:2bodymodel}), from which the parameter $\beta_{ee}$ can be determined as
\begin{equation}
\beta_{ee} = 1.6 (0.8)\times 10^{-11} \; \mathrm{cm}^3/\mathrm{s}, 
\end{equation}
to which we attribute a conservative error due to the several assumptions in our theoretical model. The blue line in figure \ref{fig:ee_decay} is the result of the fit, converted back to atom number following the scaling of equation \ref{eq:scalingnN}.


\section{Conclusions}

In conclusion, we have performed high-resolution spectroscopy of a Mott insulator of ultracold $^{174}$Yb bosons in a 3D optical lattice by exciting them on the ultranarrow $^1$S$_0$~$\rightarrow$~$^3$P$_0$ clock transition. The metrological character of the transition and the narrow spectroscopic signals that we have demonstrated allow for the characterization of the Mott insulator state and for the determination of the lattice sites occupancies. 

Our spectroscopic resolution allowed us to precisely determine the scattering lengths for $e-g$ and $e-e$ collisions in ultracold $^{174}$Yb atoms, that were previously unknown. These results are important in quantum information and quantum simulation applications, as well as for the development of optical lattice clocks based on bosonic isotopes of two-electron atoms, where the simpler internal structure (due to the absence of a nuclear spin) could provide advantages over the more commonly used fermionic isotopes. 

We have also detected the effect of inelastic collisions involving the atoms in the $e$ state. While the observed lifetimes appear to be severely limited by inelastic $e-e$ losses, no inelastic collisions in the $e-g$ channel could be observed on the timescale and with the sensitivity of our experiment. This system offers rich possibilities for quantum simulation, for instance for the investigation of two-component Bose-Hubbard models with different mobility of the species, e.g. for the study of impurity physics (a state-dependent lattice can be used to freeze the motion of the $e$ atoms in such a way to inhibit inelastic losses), or for the realization of dissipative lattice models when $e-e$ losses are taken into account.

On a more general perspective, the spectroscopic approach that we have employed could be extended in future works to the use of the same ultranarrow clock transition to probe excitation spectra of more complex quantum many-body states of either bosonic or fermionic atoms with metrological accuracy.

{\it Final note:} During the completion of the work, we became aware of very similar measurements performed by the BEC group at LKB \cite{gerbierdraft}, confirming our experimental findings.


\ack 
We acknowledge insightful discussions with Fabrice Gerbier and Carlo Sias. Special acknowledgments to the LENS
Quantum Gases group. This work has been supported by ERC Consolidator Grant TOPSIM, INFN project FISH, and MIUR PRIN 2015C5SEJJ. We thank the Galileo Galilei Institute for Theoretical Physics for the hospitality and INFN for partial support during the completion of this work.

\end{document}